# Scholarly journal publishing in transition– from restricted to open access

Bo-Christer Björk,
Hanken School of Economics,
Helsinki, Finland

## Abstract:

While the business models used in most segments of the media industry have been profoundly changed by the Internet surprisingly little has changed in the publishing of scholarly peer reviewed journals. Electronic delivery has become the norm, but the same publishers as before are still dominating the market, selling content to subscribers. This article asks the question why Open Access (OA) to the output of mainly publicly funded research hasn't yet become the mainstream business model. OA implies a reversal of revenue logic from readers paying for content to authors paying for dissemination via universal free access. The current situation is analyzed using Porter's five forces model. The analysis demonstrates a lack of competitive pressure in this industry, leading to so high profit levels of the leading publishers that they have not yet felt a strong need to change the way they operate. OA funded by article publishing charges (APCs) might nevertheless start rapidly becoming more common. The driving force currently consists of the public research funders and administrations in Europe, which are pushing for OA by starting dedicated funds for paying the APCs of authors from the respective countries.  This has in turn lead to a situation in which publishers have introduced "big deals" involving the bundling of (a) subscription to all their journals, (b) APCs for their hybrid journals and (c) in the future also APCs to their full OA journals. This appears to be a relatively risk free strategy for the publishers in question to retain their dominance of the market and high profit levels also in the future.

# Introduction

Many media industries have in the past two decades been profoundly affected by the rapid evolution of the Internet. New ways of disseminating knowledge and entertainment have been discovered, new business models have been invented and old business models have lost ground or perished. The most dramatic example is offered by encyclopedias, where Wikipedia in just a few years has more or less destroyed a two hundred and fifty years old branch of publishing. In this media landscape one of the least affected areas has been the publishing of scholarly peer reviewed journals, a global business with a turnover of roughly 10 billion dollars. This market segment is least affected in the sense that the same companies as before not only still dominate the market, but also still enjoy very high levels of profit, comparable to companies like Apple and Google. While there has been a change to mainly digital distribution of the content, the fundamental revenue model of selling content to subscribers is still dominating. When so many other areas of commerce have been rapidly transformed by the disruptive business models that the Internet offers, why have the publishers of academic peer reviewed journals been relatively unaffected. Why in particular have they been so slow to adopt the Open Access model, which would be in the best interest of just about every other stakeholder in the process? These are the major questions asked in this article.

## Scientific publications as a public good?

Free access to content, for which there is strong pressure in many media sectors, is particularly justified for scientific publications, because these can be considered a public good, in the same way as for instance laws or government reports, for which the Internet provides an excellent vehicle for free access. Typical for such public goods is that consumption of them by one individual does not reduce the amount available for other individuals. Intertwined with this is a moral argument, it simply seems wrong that scientific knowledge produced primarily by public taxpayer money should be kept locked up behind paywalls, especially since the current subscription access is in the end to a large extent paid by the same taxpayer funding.

Looking at it from the perspective of almost all stakeholders, it is a paradox that the results of the research activities of globally seven million researchers receiving mostly public funding of around 1,000 billion USD annually (Royal Society 2011), should be mostly hidden behind pay-walls. These paywalls protect the less than 10 billion USD business of one group of intermediaries, who many criticize for adding little of value, since both the articles and the work of subjecting them to review before they are published are provided for free by academics. According to this view academia as collective gives away the content for free and then, those who can afford it, buy it back, when it should be in the public domain.

Open access is not only a more cost-efficient dissemination method compared to the current subscription model, it also increases the societal impact of the research (Houghton et al 2009). The logical follow-up question is, if there are currently forces at play that might rapidly alter the situation?

# Types of Open Access

The term Open Access (OA) is usually restricted to the free availability over the Internet of scientific publications, i.e. peer reviewed journals, books and reports as well as data (Suber 2012). In other media fields the term Open Content is more common and Open Course Ware has been used to describe freely available educational material.  The rest of this article will focus narrowly on Open Access to peer reviewed journals and their articles only. The openness is technically determined by the availability to anybody with Internet access to the full text content for reading on the screen or downloading, with no barriers whatsoever like the need for payments or registration.

Many OA proponents do not accept the technical free availability alone as full Open access but also require that the articles are made available under a license explicitly enabling reuse, text mining etc of the material (i.e. a Creative Commons License). This is called libre OA in contrast to gratis OA (Suber 2008). Another requirement is immediate availability upon publication. Nevertheless delayed OA is also quite common, especially many top journals in particular in biomedicine are made freely available after a delay of 6 to 12 months (Laakso and Björk 2013).

Open access at the point of publication is often called gold OA, in contrast to green OA, which means that the authors or some third party make freely available a copy of usually the author's manuscript from some stage of the peer review process (Harnad et al 2004). Typical sites for this are home pages, institutional repositories, subject-based repositories like Pubmed Central and increasingly academic social media like ResearchGate. Green OA offers substitute access to part of the material locked behind pay walls, but it is patchy and often with available only after a delay (Eisen 2015)(Björk et al 2014). Green OA is restricted by the publishers' copyright policies, and in the past couple of years many publishers have imposed longer embargo periods, before the posting of green OA copies is allowed.

There are several alternative ways in which gold OA can be financed. In the late 1990s OA journals were mainly founded by enthusiastic scholars, using home made software and the web site of the editor's university (Björk et al 2016). There was essentially no money involved since everything, including copy editing, was done by volunteers. Starting around the year 2000 this was followed by a wave of mainly society journals or journals published by universities, which when they started producing a digital version decided to make this free. The digital version was usually subsidized by income from the print subscriptions or membership fees. This model has worked particularly well in Latin America, where journal portals like SciELO.org  have offered free technical infrastructure for hundreds of OA journals (Packer 2009).

Advertising, which in many media industries is a central revenue source enabling free content, has never been very important in the scholarly journal setting, except possibly some journals with big subscription bases with practitioners outside academia. Instead requiring the author or his institution to pay a fee for the publishing and dissemination services, was the business model adopted by new specialized OA publishers, which entered the market starting around 2001-2002. Such fees are usually abbreviated APCs, which can be interpreted as article processing charges, or article publishing charges. Although some of these publishers have become quite successful, their overall share of the number of articles published is still low, under 10 % of ISI indexed articles (extrapolated from Laakso and Björk 2012, Redhead 2015). The big subscription publishers have continued their business more or less as usual, although they have started experimenting with OA on a small scale.

A short history of the business of scholarly journal publishing

In order to understand the current situation from a business viewpoint, it is useful to look at how scholarly journal publishing has evolved throughout a number of time-periods (Tenopir and King 2000). Between 1665 and 1945 journal publishing was mainly a non-commercial activity carried out as a service by scholarly societies, a central part of their mission. Journals often had broad scopes coinciding with the areas of the societies in question (e.g Philosophical Transactions of the Royal Society, The Journal of the American Medical Association). Individual subscriptions to the paper journals were cheap, in many cases included in the membership fee.

From 1945 to 1995 publishing grew rapidly both in the number of articles published and journals. After the Second World War governments all over the world increased financing of R&D, as well as higher education, and the number of academics grew rapidly. This resulted in a strong demand for more outlets, and commercial scientific publishers rapidly increased their market share. Firstly, they often attracted authors by waiving page charges, which many society journals had. Secondly they were more agile in launching journals for new emerging niche areas, which attracted authors in the areas in question.

In the mid nineties, the World Wide Web emerged as the disruptive technology that since then has revolutionized so many markets. Around the millennium shift most of the big commercial and society publishers developed web-based platforms both for publishing parallel electronic versions of their journals and for managing the workflow of the peer review process. In terms of business this offered the opportunity both of bundling and unbundling. "Big deals" between the publisher and individual universities or consortia, covering all or big parts of a publisher's journal portfolio, became the dominant dissemination mechanism (Frazier 2001). The big deals were a win-win proposition both for the seller and buyer. In particular participating universities could offer access to vastly bigger numbers of titles than before to their faculty and students. Due to the lack of any useful usage statistics to base pricing on, the strategy of a publisher was usually to offer a given university a big deal covering several times more titles than before, for a slight mark-up compared to what they had paid earlier (Edlin and Rubinfeld 2004). Once the universities had made the first of such contracts, a strong lock-in situation was created, which has enabled the publishers to continue with yearly price increases that not only exceed inflation, but also the growth in library budgets.

In addition it was technically rather easy for the publishers to also unbundle journals and to set up the sale of individual articles (pay per view). However, this has not become very popular, perhaps because the readers using this option would mostly have had to find the funding for this from departmental or project funds, in contrast to subscription access handled centrally by the library, which for them for all practical purposes has the characteristic of a free good.

Since around 2002 new independent professional OA-only publishers have emerged on the market. Mostly these are purely commercial, with the exception of Public Library of Science, and all fund the publishing by APCs. While most of these strive to publish scientifically serious journals, the ease of setting up electronic-only journals has also enabled so-called "predatory" publishers, who for payment will rapidly publish manuscripts with little or no peer review, to enter the market (Bohannon 2013). Such publishers and in particular the amount of spam email they send academics, have to some extent tainted the reputation of serious OA publishers and journals.

The big commercial and society publishers have continued to work mainly in the subscription market but have started to experiment with so-called hybrid open access (Björk 2012).  In this model authors pay APCs to the publisher in order to provide unrestricted access to their articles amidst content available only via subscription. The model was pioneered on a bigger scale by Springer in 2004, which also set the level of 3,000 USD as the de-facto price standard. The number of journals offering this option has grown rapidly to around 11,000 (RIN 2015) and now encompasses the vast majority of journals from the major publishers. It has so far not been particularly popular among authors if they have had to pay the charges from their own, departmental or project sources. Moreover, there have been charges against the publishers for "double dipping" in terms of charging twice for the same content.  On a small scale the major publishers have also started to experiment with full OA journals, either by converting journals or by starting new ones. In particular many leading publishers have started so-called mega journals in the wake of the phenomenally successful PLOS ONE (Björk 2015). Mega journals are OA journals with a new type of peer review which is limited to checking for scientific soundness only, with typical acceptance rates of 50-70 %.

Despite the fact that there are more than 11,000 journals indexed in the Directory of Open Access Journals (DOAJ), and that these publish around half a million articles per year, the share of OA-articles in the journals indexed in the general journal indexes Web of Science or Scopus is currently around 15 %. Of these roughly half are in journals that charge APCs (Laakso and Björk 2012). The share of APCs of the total journal revenue market is still estimated to be as low as around 389 million USD in 2016 (Auclair 2015). Recent studies have, depending on the methodology used, found slightly different shares of green OA (Björk et al 2014, RIN 2015), but between 30 and 40 % of all articles can be found either in Gold OA journals, as hybrid OA articles or as green OA versions, within a year from publication.  The share of articles that could be self-archived within a year is much higher than the current level of 15-20 % (Laakso 2014), but despite a decade of universities setting up institutional repositories, in some cases also issuing OA-mandates, scholars are quite unwilling to invest the minimal time and effort needed. This partial failure is one of the reasons central research funders, in particular in Europe, have started to develop new funding instruments that could accelerate the OA development via gold and hybrid OA.

## Analysis of the current competitive situation

Why is it then that the transformation towards full gold OA has been so slow? This question can in fact be broken down into two parts. Firstly, are current leading publishers, both commercial and society, facing pressure to seek new business and revenue models? Secondly why have they been so slow in starting to adopt the business model of Open Access funded by APCs. If they wish to change business model, there are other options of which they have actively taken the bundled e-license into use as well as on a smaller scale pay-per-view. Other options could have been using third party intermediaries collecting content from several big publishers. Such intermediaries have become common in other industries, for instance iTunes, Netflix, Spotify, but not so much in scientific publishing. Examples include EBSCO and ABI/INFORM, but their services are often reduced, for instance access to some journals may entail a delay of one year. Hence most universities have made big deals directly with the big publishers (Science Direct, Springer Link) and such intermediaries have never acquired a big market share. The move to the big deals has in fact reduced the need for the big publishers to use third parties.

A useful analytical framework for understanding this state of affairs is Michael Porter's five forces model for understanding the level of competition within a particular branch of industry. This framework has been standard textbook material for decades in strategy courses at business schools (Porter 1980), but there is only one previous example of an attempt to use it for looking at scholarly publishing (McGuigan and Russell 2008). The five forces that Porter claims define the overall competitive situation are: in the center the industry rivalry, and coming from four different sides the bargaining power of suppliers, the bargaining power of buyers, the threat of new entrants and the threat of substitutes (Porter 2008). If rivalry between existing companies is fierce, suppliers and buyers have strong bargaining power. If there are threats from new entrants and substitutes, then profit margins in general will be low. At the other extremes are a few highly lucrative industries, where most of these conditions don't apply.

Applied to scholarly journal publishing the rivalry between major publishers (half a dozen commercial publishers, a few big society publishers and two big university presses) is weak. All control portfolios of usually well established journals, and in particular the four biggest commercial publishers are highly diversified across all fields of science. They don't compete on price nor do they try to get customers from each other, since almost all universities are forced to buy from all of them. The competition for market share is rather via buying up smaller publishers in mergers, the latest being the acquisition of Nature Publishing Group by Springer, as well as a competition for contracts to publish society journals on their behalf. More and more the situation has tended towards an oligopoly.

The main suppliers of these journals and publishers are the authors who provide their articles for free, or rather trade them for the publishing services and in particular the branding and reputation that these journals provide. Publishing in the "right" journals in the discipline specific pecking order is absolutely crucial, in particular for younger academics competing fiercely for positions and grants. The global trend of the last decade to primarily use publishing in highly ranked journals as the metric of academic contribution has reinforced the position of the incumbent publishers.

Additional suppliers are the many unpaid editors and reviewers who work for the journals. These also receive indirect compensation, but like the authors, in the form of building up their social capital within the networks of academics in their fields. Since the currency that authors, editors and reviewers seek is mainly in the form of academic credit and reputation, their power over the revenues of publishers is virtually nil. Hence it is not surprising that any unease from academics over the scale of those revenues and over high subscription prices has been in practical terms muted, and the few boycotts of specific journals or publishers have failed

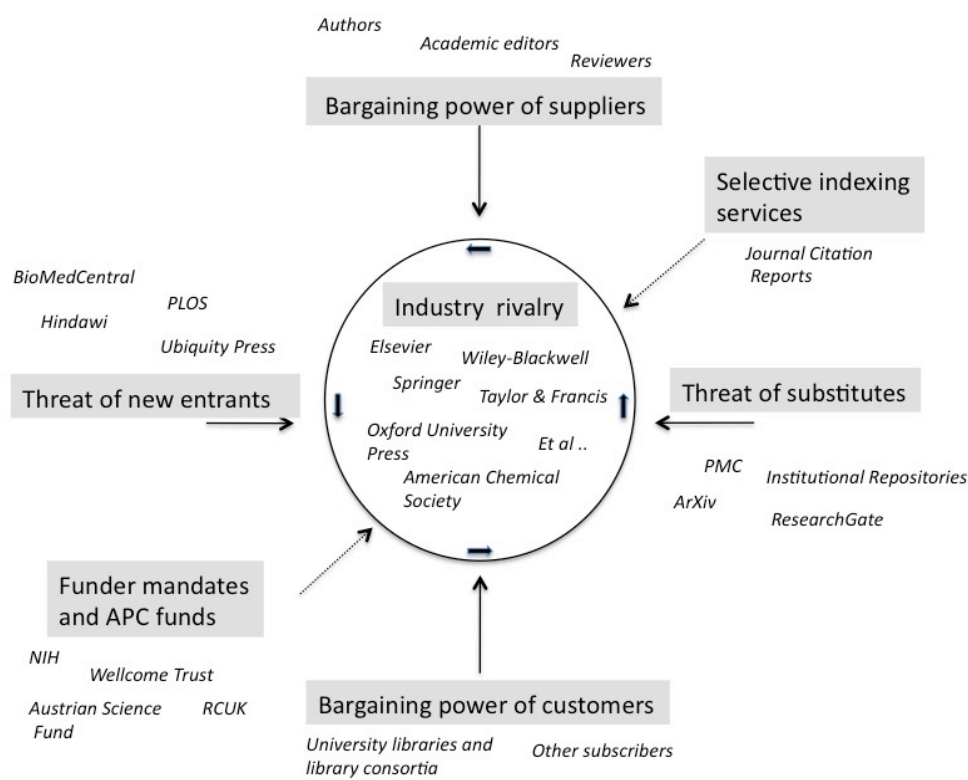

*Figure 1. Porter's five forces model adapted for analyzing the competitive situation in the scholarly journal publishing business. Two additional forces have been added, Selective indexing services and Funder mandates and APC funds.*

On the customer side university libraries and their consortia contribute around 70 % of the total subscription revenue of the publishers (Ware and Mabe 2015). The transition to online provision of journals was swiftly followed by the 'big deals' under which many universities secured access to many more journals from the major publishers, in return for a relatively small increase in the amount charged for subscriptions. Since then prices for the big deals have continued to increase faster than inflation. Library budgets have not substantially grown and hence a side-effect has been that a few big deals occupy a larger and larger share of the libraries serials budgets, squeezing out individual subscriptions and e-licenses with smaller publishers. It has also made it exceedingly difficult for new subscription publishers to enter the market. This was the major reason that Egyptian publisher Hindawi in 2005-2006 made a strategic decision to go over to OA publishing, which in hindsight turned out to be successful (Peters 2007).

This market has in fact evolved into one in which the publisher tries to price its subscriptions according to each client's ability and willingness to pay, not according to average or marginal cost for the publisher. Non-disclosure agreements make information asymmetrical, in the sense that the publishers have full information across their clients, but clients don't know exactly what other similar universities are paying. Despite occasional threats not to sign agreements, and often lengthy negotiations at renewal time of these typically three-year agreements, universities find themselves in a strong lock-in situation with all the major publishers. All in all the bargaining power of the buyers is low. And it is important to note that since it is universities and their libraries at institutional level who pay the subscriptions, academics as both authors and readers are largely insulated from any consideration as to the costs of the journals they use.

The threat from new entrants to the traditional subscription market is virtually non-existent, since it takes decades to build up a portfolio of reputable journals. The threat is further diminished by the restrictive inclusion policies of the web of science in its index, from which the journals impact factors are calculated. The web of science yearly only accepts a small percentage of new journals applying for inclusion. And in many countries and universities there are journal ranking lists which explicitly require inclusion in that index and the resulting impact factor measuring relative citedness, as a minimum criterion (this is for instance the case in the university of this author) It is for this reason that an additional force – selective indexing services - was added to the model in Figure 1. The threat from new pure OA journal publishers is bigger, although these too are to some extent hindered by the barrier of getting into the journal indexing services. And where new OA publishers have succeeded, the major traditional publishers have taken measures to counter the threats they pose. BiomedCentral was purchased by Springer in 2008, and recently most of the traditional publishers started mega-journals of their own, increasingly competing with PloS ONE (Björk 2015). All in all the threat has until now been low.

The threat from substitutes is an interesting issue. Wikipedia, Airbnb or even Uber are real substitutes for older established services or middlemen providing access to such services. But there is no substitute for a specific journal article in a prestigious scholarly journal, you have to get access to just that article or its manuscript. Individual readers can try to find a free green OA copy, but such copies can still only be found for a minority of articles, and the availability is highly random. So far, green OA has not threatened the profits of the leading subscription publishers. University libraries, who are the main source of subscription revenue, cannot rely solely on directing their users to the haphazard availability of green copies, but need to buy the standard access from the publishers. But if there were signs that the threat to their subscription revenues was becoming real, publishers would start to further tighten the embargo rules of green OA (Kingsley 2014) and probably start legal actions against academic social networks (Robinson 2014).

Already this superficial analysis would indicate an industry, where all the five forces (plus the added sixth force of the selective indexing services,) would work in the same direction to result in an extremely low level of competition, which would manifest itself as a high level of profitability.

The operating profit levels, defined as the share of profits before taxes of the total revenues, were all in the range 32-42 % in 2010-2012 for leading publishers Elsevier, Springer, Wiley-Blackwell and Taylor and Francis (Morrison 2012, the Economist 2013). Another industry characteristic of interest is the level of concentration. Larivière et al (2015) have systematically studied the shares of the five biggest publishers of articles indexed in Web of Science. In STM the share increased from 20 % in 1973 to 30 % in 1996 and is currently 53 % (with the three biggest publishes having a 49 % share).

Against this background it is very easy to understand why the established publishers have been very slow to start experimenting with open access. Why should they, with their current high profit levels, change their business plans? Going into the OA market would be very risky, and their current sales revenues are protected against rapid fluctuations by license agreements spanning several years as well as the fact that university libraries are in a lock-in situation. The APC levels paid by important research funders mainly to the independent OA publishers have so far been much lower on average (Kiley 2015, Schimmer et al 2015), than the levels publishers claim they would need to cover their costs and clearly lower than the current charges of hybrid OA journals.

## A Scenario for the future

So far the two major scenarios for moving to full OA have been a coexistence of subscription journals and green self-archived copies achieving OA, alternatively full OA journals, in particular from new publishers, gradually taking over the market.
But are there currently developments that may disrupt the balance of forces and speed up the move towards full gold OA from the current 1-2 % yearly increase in its share of global article volume? A seventh force, which potentially might disrupt the balance in favor of new entrants, consists of the OA mandates and accompanying APC funds of major research funders, and is also indicated in figure 1. Early indications of this can be seen in the UK, where such measures have resulted in rates of increase in OA take up much higher than the world average, with especially high rates of increase in take up of hybrid OA (RIN 2015). This force, as well as how the major subscription publishers might adapt their strategies to prevail, is discussed below.
Until now the major publishers have not seriously started to use the open access business model for whole journals, but have experimented with it only at a small scale. Instead Springer started hybrid OA for most of their titles already in 2004, and between 2009-2013 the number of hybrid journals grew rapidly from around 2,000 to around 8,000. Today the vast majority of journals from all the big subscription publishers offer this option. Until around 2014, the relatively uniform and high price level of around 3,000 USD resulted in generally very low uptake levels of around 2 % of eligible articles. Subscription publishers have converted to full OA only a small number of their journals. Oxford University Press was among the first in 2005 by converting their flagship journals Nucleic Acids Research as part of a systematic experiment also involving hybrid journals (Bird 2010). In the past couple of years the number of conversions has increased, often involving journals that a commercial publisher has published on behalf of scientific societies (Solomon, et al 2016)). Some newly- started OA journals have been megajournals, which benefit a lot from the publishers trademark, and which can tap into a pool of manuscripts rejected from the more selective journals of the publishers in question. In addition a useful strategy is also the acquisition of successful OA publishers (ie. Springer bought BMC in 2008).
But the situation may be about to start changing. The main agents of change are the major research funders in Europe. Both the European Union and major national researcher funders as well as ministries of education have made achieving Open Access an important goal. Since the UK Finch report (2012) there has been a shift of focus from green OA to gold OA. The key ingredient in such policies has been a growing awareness that green OA (in particular via institutional repositories) is not succeeding to the degree hoped for, as well as of the need to provide earmarked funds for APCs. Such funds have thus been created in particular in the UK via the research councils as well as Wellcome Trust, but also in other European countries (Fransvåg 2014, Austrian Science Fund 2014). Administratively the favorite arrangement seems to be that the central funders require universities to set up their own central funds, which administer all the payments, and then get reimbursed from the central funds.

The exact rules for what sort of APCs are reimbursed are obviously of great interest to publishers. In the early days funders like Wellcome Trust reimbursed both full OA and hybrid OA payments to 100 %, which obviously puts no pressure on the publishers to curb the prices. For this reason the Wellcome Trust together with other important European research funders commissioned a report studying alternative mechanisms to put price caps on the APCs that they automatically refund (Björk and Solomon 2014).

In order to counter the charges of "double dipping" (Kingsley 2014), i.e. charging twice for the same articles, some publishers have created different sorts of offsetting deals, which provide discounts on APCs, or rebates on subscription fees based for the articles in hybrid journals from institutions which sign such agreement (Estelle 2014, Geisenheimer 2014).

The latest development is that some big publishers have started to offer new sorts of consortial e-licenses which cover not only subscription but also the hybrid payments (Poynder 2014, Austrian Science Fund 2015). If this strategy is widely adopted it could, in the longer term, lead to a mass conversion of all the journals of major publishers to full OA (Shieber 2015). The strategy would be a direct continuation of the transition from individual paper journal subscriptions to bundled e-licenses.

In a similar fashion the currently emerging big deals are probably not much more expensive than the current subscription licenses, but include free OA, at least hybrid OA, for all authors of the institutions in question in all the journals of the publishers in questions. Such deals were already tested by Springer in 2007-2010 with, for instance, the University of California Libraries, The Dutch University Libraries Consortium and the Max Planck Society (Albandes 2009).

What could make such deals attractive, or at least acceptable both for seller and buyers? From the viewpoint of the universities the important issue would be that they wouldn't have to increase their budgets more than marginally for journal acquisitions and APCs combined, something Pinfield et al (2015) call "total cost of publication". The first movers to sign long-term agreements may also be getting advantages (Crotty 2016). In particular it would seem to be attractive for top universities to sign such deals, since estimates have pointed out that research intensive universities in a future 100% OA world with APC fees would end up paying much more compared to what their share of subscription fees is (Walters 2007). It will be more problematic for less-research-intensive universities, which could envisage a role as free-riders in the APC era, having to pay much less to secure access to journals and their contents. But the bigger the big deals tend to become (eg all UK or Dutch universities) the more differences between universities with different authorship/readership profiles would tend to average out. All in all it seems pretty obvious though that few universities would be willing to sign agreements where they pay substantially more than currently, even if they might recoup part of the money from research contracts which include provision to meet the costs of APCs.

The publishers again probably value the long-term stability of such deals, along with the prospect, after a certain critical mass of hybrid OA articles has been achieved, of a gradual relatively risk-free transition to full OA, as described by Prosser (2003). The major subscription publishers also have a big competitive advantage compared to the competing full-OA publishers, since they have already carved out their slices of library budgets, and have all the negotiation relationships and routines in places.

While the earlier big deals bridged the transition between print and electronic journals, this new type bridges three evolutionary stages, subscription, hybrid and full OA. Even though the deals currently may only include subscriptions and free APCs for hybrid OA in the same journals, it is logical to ask what will happen once the hybrid shares in these journals start to approach a critical mass of say a third or more, and in a more even pattern than before? There would be mounting pressure on the publishers in question to start lowering the subscription fees to all other subscribers, so at some point they would have to start flipping the journals altogether to OA. Since their agreements already include "free" open access publishing in all the journals that were hybrid at the time the agreements started, it would be difficult for them to start charging extra APCs for such journals they converted to full OA during the time period of the agreements (which are often for three years). And also difficult to justify raises for those journals in later deals with the universities in question.

Although publishers would like to recoup as much in APCs in the future as they currently receive in subscription fees, this is perhaps not possible via the big deals. Springer at least seems to be trying by first estimating the publication output of the institution in question in its hybrid journals and then calculating the total price at the standard 3,000 USD hybrid rate and then using these payments to largely offset the subscription price (Ritt 2015). Since the details of these agreements are usually protected by non-disclosure clauses, it is difficult to know the exact resulting price level in the first of these deals.

Importantly, if the publishers were not to work via big deals, the conversion of journals would have to be done more individually, involving bigger risks. In principle it would be possible for them to get more revenues by selling individual APCs at high enough prices. But that would mean that they will face much tougher competition from new entrants to the market, which usually have lower APCs, for comparable quality journals.

What we also might see is a split of the APC- funded OA market into two parts. One in which the APC level of individual journals is an important competitive factor and authors may have to find the money to pay the APCs from their projects, departmental funds or even their own money. This submarket would include most APC-charging journals not included in the big deals of the major publishers. In that setting they may become very sensitive to the price and particularly the perceived value compared to the price (Björk and Solomon 2015). Here price competition from new low-cost publishers like Ubiquity Press can be important. The second market would be the one covered by the big deals encompassing subscription, hybrid and converted OA journals combined. In that market authors are totally insulated from the APC and publish in the same journals as before, based on other considerations. The situation is pretty much the same as currently with regard to subscription journals: academics both as authors and readers would remain largely insulated from any consideration relating to costs.

## Conclusions

Much has been written about the current state of affairs in scholarly journal publishing, about subscription prices rising faster than inflation (often referred to as the serials crises), and about the potential benefits of Open Access. Using the lens of Porter's five forces model helps to further focus in on the current competitive situation (or rather lack of competition) in order to understand better why the leading publishers have been slow to change their revenue model from selling content to selling OA dissemination services.

If the OA revolution is to come about via newly founded OA journals replacing old established ones, or major publishers converting individual journals one by by, the current slow evolution trend of 1-2 % market share increase per year will continue. The main reason is that authors choose where to submit largely based on very stable journal rankings and the effects of the journal impact factor "institution", which shelters those journals included in the citation indices from competition from new entrants.

It is the firm belief of this author, that the only way a rapid transition to full gold OA is possible, is via a massive conversion to OA of the portfolios of the major publishers - including the leading journals in which academics actually publish - via big deals that bridge the transition from subscription to OA. And given the currently comfortable situation of the major subscription publishers, they will only be willing to do this if they can do it in the relatively risk-free way of augmented big deals, that guarantees them close to the same revenues as today. Obviously the scenario outlined in this paper is highly speculative; on the other hand ignoring the possibility of this scenario in a discussion about the future of scholarly journal publishing would be short-sighted.

## Acknowledgements:

Michael Jubb provided very useful comments to a draft version of this manuscript.